\documentclass[final,3p,times,twocolumn]{elsarticle}
\usepackage{xcolor}
\usepackage{subfigure}
\usepackage{hyperref}

\usepackage{amssymb}
\usepackage{amsmath}

\begin{document}

\begin{frontmatter}

\title{QUANTUM ESPRESSO implementation of the RPA-based functional}
\author{Angel Rosado}
\author{Mario Benites}
\author{Efstratios Manousakis}
\affiliation{
    organization={Department of Physics, Florida State University},
    city={Tallahassee},
    state={FL},
    postcode={32306-4350} }

\begin{abstract}
  We detail our implementation of the random-phase-approximation based functional (RPAF) derived in Ref.~\cite{benites2024} for the QUANTUM ESPRESSO (QE) package.
  We also make available in the {\it Computer Physics Communications} library the source files which are required in order to apply this functional within QE. We also provide the corresponding RPAF projector
  augmented wave (PAW) and
  ultrasolf pseudopotentials for most elements.
  Lastly, we benchmark the performance of the RPAF by calculating the equilibrium lattice constant and bulk modulus of a set of the same 60 crystals used by other authors to benchmark other functionals for both PAW and ultrasoft pseudopotentials. We find that the RPAF performs better overall as compared to the other most popular functionals.  
\vskip 0.3 in
  
\noindent \textbf{PROGRAM SUMMARY}

\begin{small}
\noindent
{\em Program Title: Implementation of RPAF functional in QUANTUM ESPRESSO} \\
{\em Developer's repository link:} \href{https://dftrpa.github.io}{https://dftrpa.github.io} \\
{\em Licensing provisions: GPLv3}  \\
{\em Programming language: Fortran 90} \\
{\em Nature of problem: To make the RPAF available to be used in DFT calculations}\\
{\em Solution method: Implementation of RPAF in QUANTUM ESPRESSO}\\
   \\
\end{small}
\end{abstract}
\end{frontmatter}


\section{Introduction}
\label{section:introduction}

The electronic structure has become one of the main components of characterization of materials across various scientific disciplines.
It provides not only the foundation to understand basic behavior of materials, but, in addition, the cornerstone to any further attempt to understand more complex phenomena.

While the many-body problem of the interacting electrons surrounding an array of nuclei is intractable, there is a simplified approach founded\cite{PhysRev.136.B864,PhysRev.140.A1133} on the
picture that the single-electron degrees of freedom can be considered as embedded in an effective field produced self-consistently by the collective pressence of
the rest of the electrons through their density field $n(\vec r)$. The emerging mathematical framework, known as density-functional theory (DFT), is based on the existence of a
one-to-one correspondence between the ground-state of the fully interacting system and this density field $n(\vec r)$. More importantly, there exists a material-independent (i.e., universal) functional of the density,
$E_{xc}(\{n(\vec r) \})$,  which plays the role of this effective interaction of each electron with the collective presence of all the rest of the electrons.

The universal nature of this functional allows its determination by simply focusing on the interacting  electron gas. A further approximation is
to exclude the explicit contribution of non-local effects and
restrict ourselves within the so-called local-density approximation (LDA).
However, there is no well-defined approach to determine the form
of the functional within LDA and the currently available methods are based on at least one ad-hoc step.
Recently, we used a rational generalized expansion approach to systematically derive it; we presented a random-phase-approximation based local-spin-density functional (RPAF) \cite{benites2024}, using an expansion 
in a dressed and well-behaved interaction vertex because it is renormalized to include the effects of  screening. The approach is detailed in
Ref.~\cite{benites2024} where the leading terms in a systematic expansion of the
exchange correlation functional are given. At the present time we are
also working on including the next order corrections.

However, as shown
here, the leading order outperforms the presently available most popular         LDA
functionals.
Therefore, it makes sense to make the source files and its
implementation available for general use and
the purpose of this paper is in part that, and, in addition,
to demonstrate its superior performance as compared to
other functionals.
More specifically, we present the implementation of RPAF in QUANTUM ESPRESSO (QE) \cite{QE-2017, QE-2009} and provide the source files containing the RPAF to be used in QE v7.2.

The paper is organized in the  following way. In the next Section, we summarize the complete form of the functional which includes a table with the values of its parameters as determined in Ref.~\cite{benites2024}.
In Sec.~\ref{section:how_to_use_rpaf} we outline how to
use the provided source file to reinstall QE for applying
the RPAF.  In Sec.~\ref{section:performance} we present the performance
of the RPAF and compare it to other functionals and in Sec.~\ref{section:conclusion}, we present our conclusions.

\section{The RPA based functional}
\label{section:functional}
The RPAF consists of the sum of two terms referred to as the ring-diagram series and the kite-diagram series:
\begin{equation}
    \varepsilon_{c}(r_{s}, \zeta) = 
    \varepsilon_{r}(r_{s}, \zeta) + 
    \varepsilon_{2b}(r_{s}, \zeta),
\end{equation}
where $r_{s}$ is the Wigner-Seitz radius and $\zeta = \frac{n_{\uparrow} - n_{\downarrow}}{n_{\uparrow} + n_{\downarrow}}$ is the spin-polarization with $n_{\uparrow}$ ($n_{\downarrow}$) being the density of spin up (down) electrons.
The energy per electron of the ring-diagram series is fitted to the form:
\begin{multline}
    \varepsilon_{r}(r_{s}, \zeta) = 
    \left[a_{0}(\zeta)+a_{1}(\zeta)r_{s}\right]
    \ln\left[1+\frac{a_{2}(\zeta)}{r_{s}^{2}} \right] \\ 
    + \left[b_{0}(\zeta)+b_{1}(\zeta)r_{s}\right]
    \ln\left[1+\frac{b_{2}(\zeta)}{r_{s}^{7/4}} \right],
    \label{equation:ring_fit}
\end{multline}
with the parameters given by
\begin{subequations}
\begin{align}
    a_{2}(\zeta) &= 
        a_{20} + a_{21}(\chi - 2) 
        + a_{22}\left(\frac{\ln\chi}{\chi} 
        - \frac{\ln2}{2}\right), \\
    b_{2}(\zeta) &= 
        b_{20} + b_{21}(\chi - 2) 
        + b_{22}\left(\frac{\ln\chi}{\chi} 
        - \frac{ln2}{2}\right), \\
    b_{0}(\zeta) &= 
        \frac{2c_{0}(\zeta)+c_{L}(\zeta)\ln a_{2}(\zeta)}
        {2\ln b_{2}(\zeta)-\frac{7}{4}\ln a_{2}(\zeta)},\\
    a_{0}(\zeta) &= 
        -\frac{1}{2}\left(c_{L}(\zeta) 
        + \frac{7}{4}b_{0}(\zeta)\right),\\
    b_{1}(\zeta) &= 
        \frac{d_{0}(\zeta)}{b_{2}(\zeta)},\\
    a_{1}(\zeta) &= 
        \frac{d_{1}(\zeta)}{a_{2}(\zeta)},
    \end{align}
    \label{equation:ring_parameters}
\end{subequations}

with the rest of the dependencies given by:

\begin{multline}
    c_{L}(\zeta) = \frac{1}{2\pi^{2}}\Bigg[
    2-2\ln(4)
    + \chi\chi_{+}\chi_{-} \\ \left.
    + \chi_{+}^{3}\ln\left(\frac{\chi_{+}}{\chi}\right)
    + \chi_{-}^{3}\ln\left(\frac{\chi_{-}}{\chi}\right)
    \right],
    \label{equation:ring_ln_coefficient}
\end{multline}
and
\begin{subequations}
\begin{align}
    c_{0}(\zeta) &= 
        \sum_{n=0}^{1}c_{0n}\zeta^{2n} 
        + \sum_{n=0}^{2}\bar{c}_{0n}(\chi^{n+1} - 2^{n+1}), \\
    d_{0}(\zeta) &= -0.803,
        \\
    d_{1}(\zeta) &= \sum_{n=0}^{2} d_{1n}\zeta^{2n},
\end{align}
\label{equation:extra_parameters}
\end{subequations}
where $\chi_{+} = (1+\zeta)^{1/3}$, $\chi_{-} = (1-\zeta)^{1/3}$, and $\chi = \chi_{+} + \chi_{-}$. 
The coefficients are given in Table \ref{table:coefficients}. The coefficients 
$c_{0n}$, ${{\bar c}_{0n}}$, $c_{L}$, $d_{0}$, $d_{1n}$ and the parameters $A_{0}$ and $A_{2}$ are in Rydberg energy units.

The energy of the kite-diagram series is fitted to the form:
\begin{multline}
    \varepsilon_{2b}(r_{s}, \zeta)  = \frac{A_{0}(\zeta)}{1+A_{1}(\zeta)r_{s}} +\\ A_{2}(\zeta)r_{s}\ln\left[ 1+\frac{1}{A_{3}(\zeta)r_{s}+A_{4}(\zeta)r_{s}^{3/2}} \right],
    \label{equation:kite_fit}
\end{multline}
where $A_{0}(\zeta)=0.04836$ Ry and the $\zeta$ dependence of the other parameters is given by:
\begin{equation}
\begin{aligned}
    A_{n}(\zeta) = 
    \sum_{m=0}^{2}A_{nm}\zeta^{2m}
    \quad, \quad \text{for} \quad
    \text{n=1, 2, 3, 4}
\end{aligned}
\label{equation:kite_parameters}
\end{equation}
with the coefficients in Table~\ref{table:coefficients}.

\begin{table}
    \begin{center}
    \def\arraystretch{1.3}
    \begin{tabular}{|c|c|c|c|} 
        \hline
        $n$ & 0 & 1 & 2\\
        \hline
        $a_{2n}$ & 90.76 & 192.62 & -3956.38  \\
        \hline
        $b_{2n}$ & 54.55 & 149.46 & -2070.06  \\
        \hline
        $c_{0n}$ & -0.1423 & 0.0036 & 0 \\
        \hline
        ${\bar c}_{0n}$ & 0.1971 & -0.0326 & -0.0177\\
        \hline
        $d_{1n}$ & 0.8822 & 0.1648 & 0.0432 \\
        \hline
        $A_{1n}$ & 0.10215 & -0.05028 & -0.01283  \\
        \hline
        $A_{2n}$ & -0.01382 & 0.00016 & 0.00808 \\
        \hline
        $A_{3n}$ & 0.46529 & 0.05868 & -0.32923 \\
        \hline
        $A_{4n}$ & 0.00364 & -0.00259 & -0.00021 \\
        \hline
    \end{tabular}
    \caption{Values of the parameters and coefficients used in the functional.}
    \label{table:coefficients}
    \end{center}
\end{table}

Besides the correlation energy functional, any DFT code needs the correlation potential as well. For the local part of the functional in the DFT equations, the potential takes the form
($\varepsilon_{c}=\varepsilon_{c}(r_{s},\zeta)$ for simplicity):
\begin{equation}
    v_{c}(r_{s}) = \varepsilon_{c} - \frac{1}{3}r_{s}\frac{\partial \epsilon_{c}}{\partial r_{s}}, \\
    \label{equation:potential}
\end{equation}
for spin unpolarized calculations and 
\begin{subequations}
\begin{align}
    v_{c\uparrow}(r_{s},\zeta) &=
    \varepsilon_{c} - \frac{1}{3}r_{s}\frac{\partial
    \varepsilon_{c}}{\partial r_{s}} + \frac{\partial
    \varepsilon_{c}}{\partial \zeta}(1-\zeta), \\
    v_{c\downarrow}(r_{s},\zeta) &=
    \varepsilon_{c} + \frac{1}{3}r_{s}\frac{\partial
    \varepsilon_{c}}{\partial r_{s}} - \frac{\partial
    \varepsilon_{c}}{\partial \zeta}(1+\zeta),
    \label{equation:potential_spin_polarized}
\end{align}
\end{subequations}
for spin-polarized calculations. These expressions are easy to implement for the RPAF since the functional is an analytic function of $r_{s}$ and $\zeta$.

\section{How to use RPAF in QUANTUM ESPRESSO}
\label{section:how_to_use_rpaf}
To implement the RPAF in QE, we modified the next four source files contained in the \texttt{XClib} directory of the QE directory: \texttt{qe\_dft\_list.f90}, \texttt{qe\_dft\_refs.f90}, \texttt{qe\_drivers\_lda\_lsda.f90}, and \texttt{qe\_funct\_corr\_lda\_lsda.f90}. To make RPAF available as an option, during the
installation of the QE one should
replace the above mentioned four files with those provided in the Developer's Repository (See link in the PROGRAM SUMMARY) and recompile QE with \texttt{make all}.
Once QE is re-installed, RPAF becomes available to be used under the functional name ``\texttt{RPAF}" or equivalently ``\texttt{XC-001I-015I}" in index notation. For example, for a self-consistency calculation in the input file of the \texttt{pw.x} code, RPAF can be specified using the flag \texttt{input\_dft="RPAF"} or \texttt{input\_dft="XC-001I-015I"} under the \texttt{SYSTEM} namelist. The \texttt{pw.x} code reads by default which functional to use from the pseudopotentials file, so if RPAF pseudopotentials are used (as the ones provided in this paper), the flag \texttt{input\_dft="RPAF"} is not necessary for the \texttt{pw.x} code.

Pseudopotentials for the RPAF were generated using the ld1.x code of QE and are provided in the Developer's Repository (See link in the PROGRAM SUMMARY). Two types were generated: projector augmented wave (PAW)\cite{PhysRevB.50.17953} and ultrasoft pseudopotentials \cite{PhysRevB.41.7892}. We used the pslibrary 1.0.0 library \cite{dalcorso2014} of pseudopotentials input files for the \texttt{ld1.x} to generate the RPAF pseudopotentials. The format for PAW and ultrasoft pseudopotentials provided here is the Unified Pseudopotential Format (UPF).

\section{Performance of the RPAF}
\label{section:performance}
The performance of RPAF was tested by calculating the equilibrium lattice constants and bulk moduli of a set of 60 crystals listed in Ref.~\cite{haas2009} and comparing the results to the experimental values \cite{haas2009, tran2007, brandes2013smithells, liang2021, gerward2005, kang2019}. We also compared the results to those obtained with the PW functional. Calculations were also done with the PZ functional, but the results were not included in the Figures provided next because they differed from the PW functional slightly and the difference would not be visible in the graphs.

We performed self-consistent calculations at different lattice constants with steps of 0.01 Bohr radius. The equilibrium lattice constant was found by finding the lattice constant that yielded the minimum energy. We fitted nine energy-lattice constant points, with four points smaller and four points larger than the equilibrium lattice constant, to the Birch-Murnaghan equation of state using the \texttt{ev.x} tool in QE. From this fit, we obtain the equilibrium lattice constant and bulk modulus.

\begin{figure*}[htp]
    \begin{center}
         \subfigure[]{
    \includegraphics[scale=0.4]{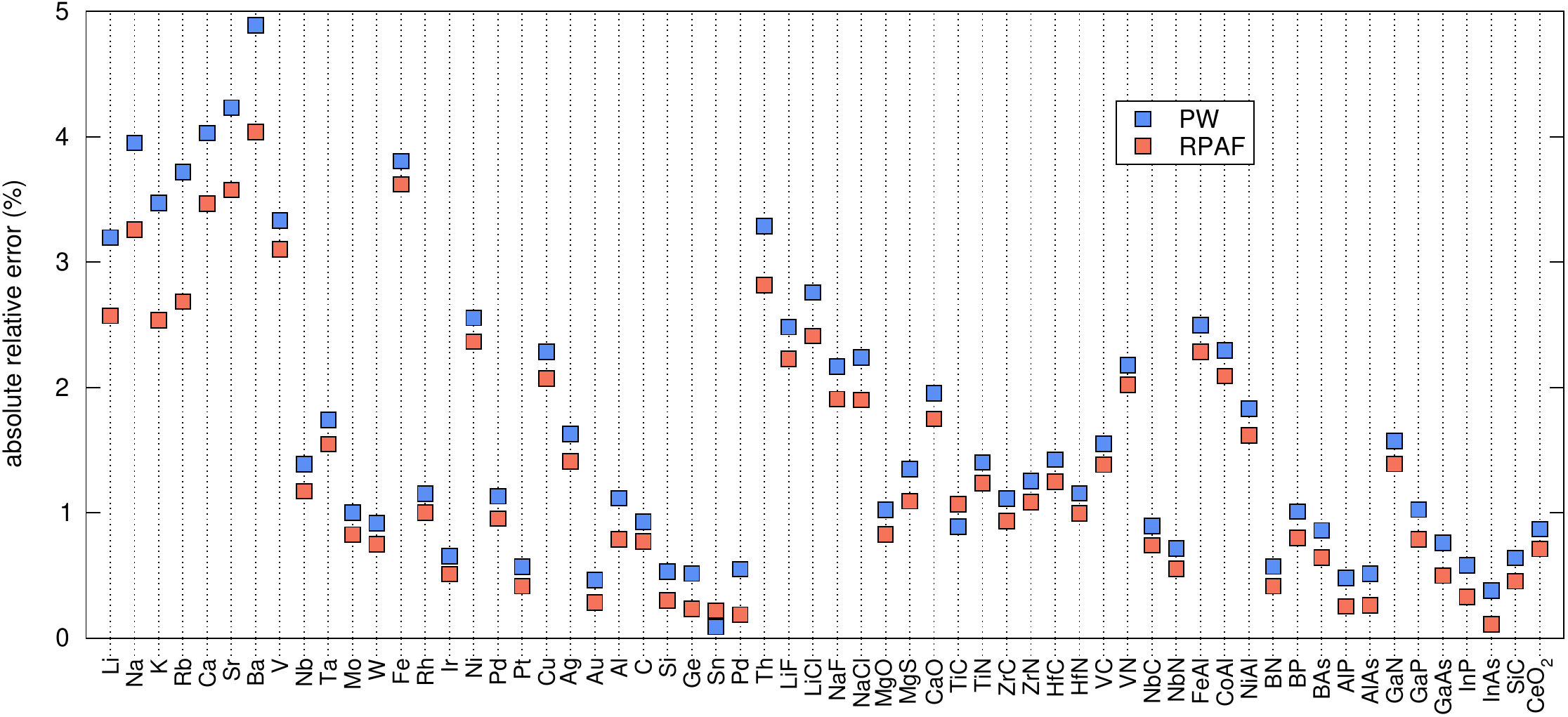}
    \label{figure:a0_paw}
         }\\
         \subfigure[]{
    \includegraphics[scale=0.4]{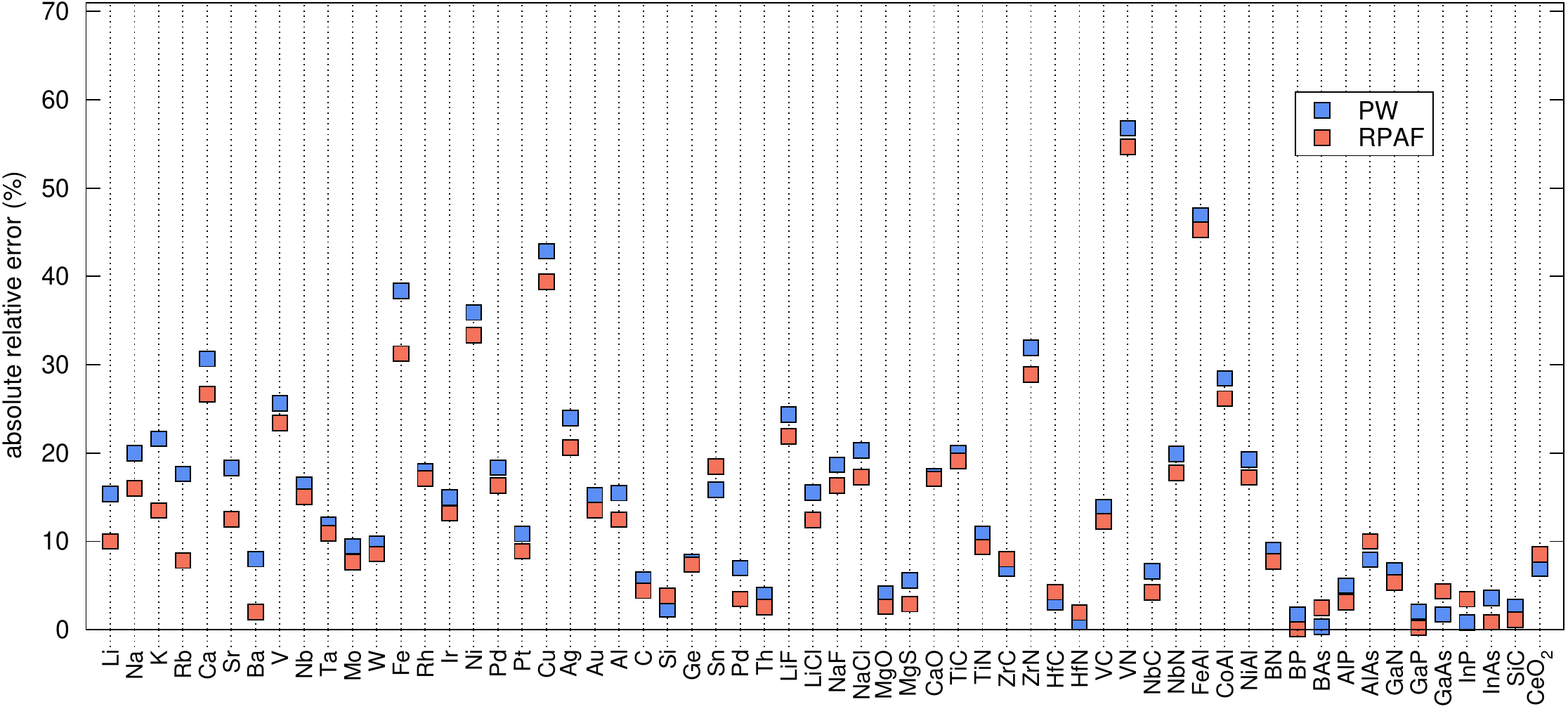}
    \label{figure:b0_paw}
         }\\
    \end{center}
    \caption{Relative errors of equilibrium lattice constants (panel (a)) and
      bulk moduli (panel (b)) of
     the 60 crystals set using the PW functional (blue) and the RPAF (red) 
     with PAW pseudopotentials.}
\end{figure*}

\begin{figure*}[htp]
    \begin{center}
         \subfigure[]{
    \includegraphics[scale=0.4]{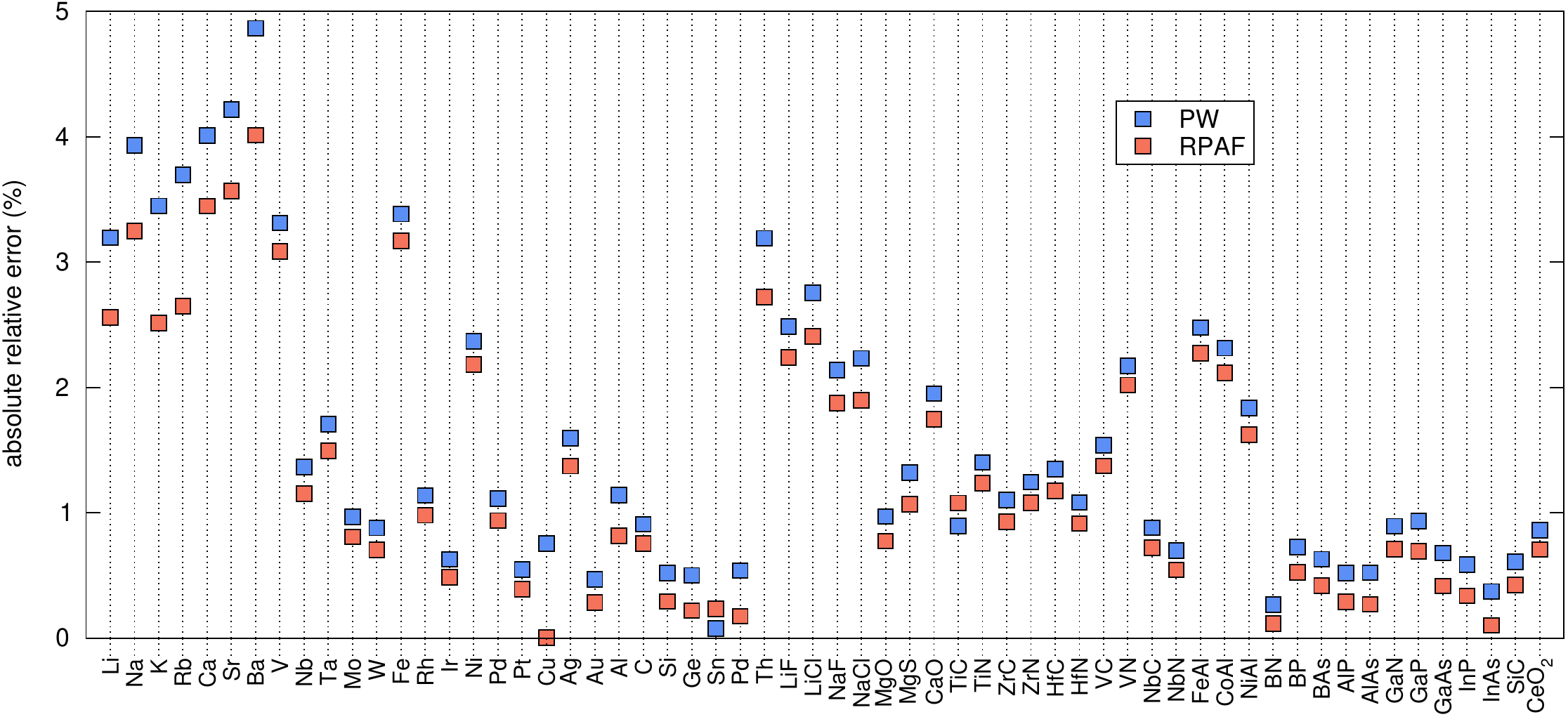}
    \label{figure:a0_us}
         }\\
         \subfigure[]{
    \includegraphics[scale=0.4]{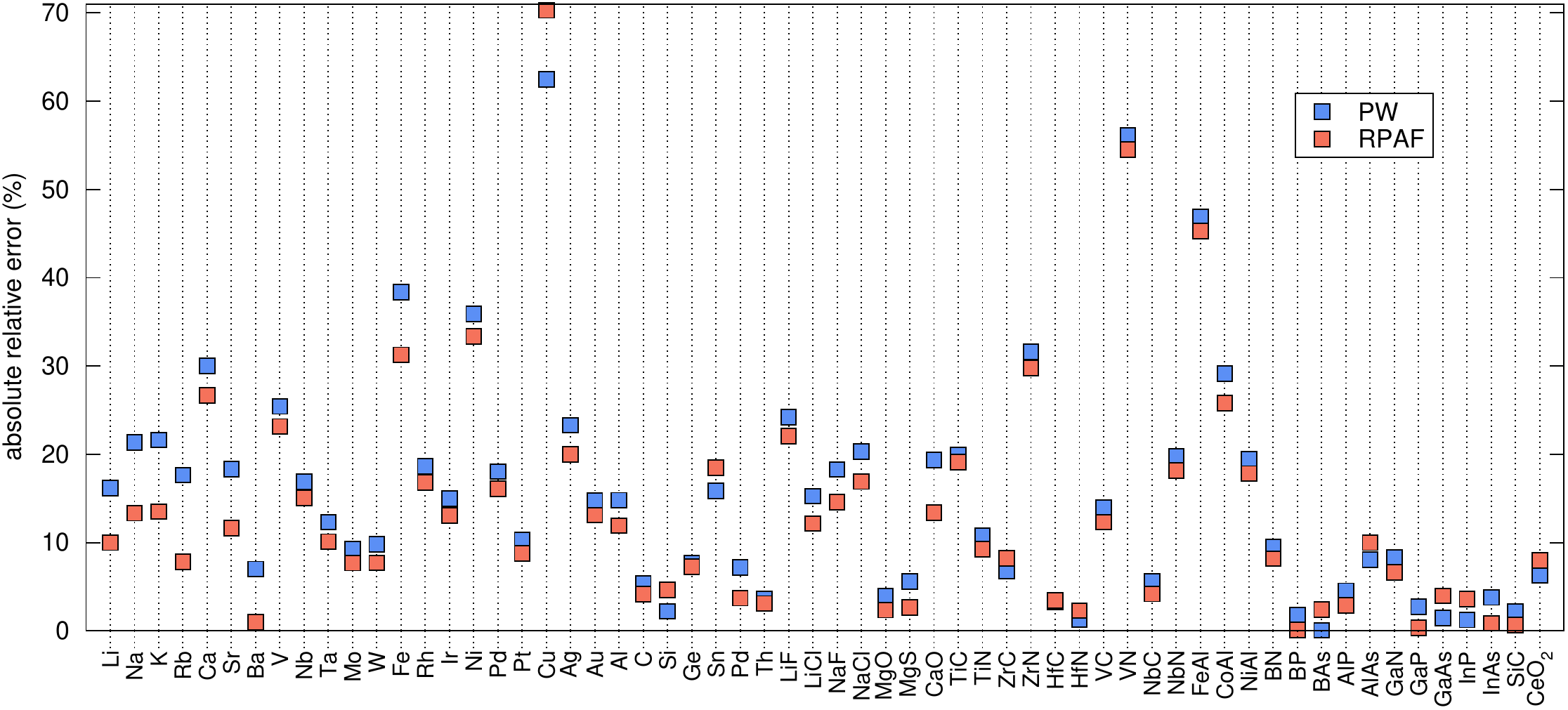}
    \label{figure:b0_us}
         }
    \end{center}
    \caption{Relative errors of equilibrium lattice constants (panel (a)) and
      bulk moduli (panel (b)) of
     the 60 crystals set using the PW functional (blue) and the RPAF (red) 
     with ultrasoft pseudopotentials.}
\end{figure*}

For the RPAF calculations, we used the pseudopotentials provided in this paper. A cutoff energy of 120 Ry for the wavefunctions, a cutoff of 1200 Ry for the charge density, and a k-grid of $12\times12\times12$ were found to be enough for convergence and to obtain a well-defined energy-lattice constant parabola. The results are shown in Fig.~\ref{figure:a0_paw} and Fig.~\ref{figure:b0_paw} with PAW pseudopotentials and in  Fig.~\ref{figure:a0_us} and Fig.~\ref{figure:b0_us} with ultrasoft pseudopotentials.

\section{Conclusion}
\label{section:conclusion}
The RPAF is a novel functional derived \cite{benites2024} by means of including the leading order in a systematic expansion approach, which can be further improved by the
inclusion of higher-order terms. However, the functional obtained from the leading order performs better than the most accurate LDA functionals available.
Therefore, it might be useful to use the present functional in future electronic structure calculations.

Therefore, one of the goals of the present paper is to make available the source files containing the RPAF for QE, and,  in addition, the PAW and ultrasoft pseudopotentials are provided with this paper. In addition,  we have explained
how to carry out a re-installation of the QE by replacing some QE source files with those provided here and how to use the RPAF. 

Furthermore, in the present paper we demonstrate that the
overall performance of the RPAF is better than that obtained with the PW (and PZ) functional for both equilibrium lattice constant and bulk modulus
using both PAW and ultrasoft pseudopotentials.

The RPAF is in the process of being implemented in the \texttt{LIBXC} library of pseudopotentials. Therefore, in the near future, the RPAF will become available to use with other
implementations of the DFT. In addition, a gradient approximation extension of the LDA, based on the RPAF functional, is our future goal.

\bibliographystyle{elsarticle-num}

\begin{thebibliography}{10}
\expandafter\ifx\csname url\endcsname\relax
  \def\url#1{\texttt{#1}}\fi
\expandafter\ifx\csname urlprefix\endcsname\relax\def\urlprefix{URL }\fi
\expandafter\ifx\csname href\endcsname\relax
  \def\href#1#2{#2} \def\path#1{#1}\fi

\bibitem{benites2024}
M.~Benites, A.~Rosado, E.~Manousakis, Accurate electron correlation energy
  functional: Expansion in the interaction renormalized by the random-phase
  approximation, Phys. Rev. B 110 (2024) 195151.
\newblock \href {https://doi.org/10.1103/PhysRevB.110.195151}
  {\path{doi:10.1103/PhysRevB.110.195151}}.

\bibitem{PhysRev.136.B864}
P.~Hohenberg, W.~Kohn, Inhomogeneous electron gas, Phys. Rev. 136 (1964)
  B864--B871.
\newblock \href {https://doi.org/10.1103/PhysRev.136.B864}
  {\path{doi:10.1103/PhysRev.136.B864}}.

\bibitem{PhysRev.140.A1133}
W.~Kohn, L.~J. Sham, Self-consistent equations including exchange and
  correlation effects, Phys. Rev. 140 (1965) A1133--A1138.
\newblock \href {https://doi.org/10.1103/PhysRev.140.A1133}
  {\path{doi:10.1103/PhysRev.140.A1133}}.

\bibitem{QE-2017}
P.~Giannozzi, O.~Andreussi, T.~Brumme, O.~Bunau, M.~B. Nardelli, M.~Calandra,
  R.~Car, C.~Cavazzoni, D.~Ceresoli, M.~Cococcioni, N.~Colonna, I.~Carnimeo,
  A.~D. Corso, S.~de~Gironcoli, P.~Delugas, R.~A.~D. Jr, A.~Ferretti,
  A.~Floris, G.~Fratesi, G.~Fugallo, R.~Gebauer, U.~Gerstmann, F.~Giustino,
  T.~Gorni, J.~Jia, M.~Kawamura, H.-Y. Ko, A.~Kokalj, E.~Küçükbenli,
  M.~Lazzeri, M.~Marsili, N.~Marzari, F.~Mauri, N.~L. Nguyen, H.-V. Nguyen,
  A.~O. de-la Roza, L.~Paulatto, S.~Poncé, D.~Rocca, R.~Sabatini, B.~Santra,
  M.~Schlipf, A.~P. Seitsonen, A.~Smogunov, I.~Timrov, T.~Thonhauser, P.~Umari,
  N.~Vast, X.~Wu, S.~Baroni, Advanced capabilities for materials modelling with
  quantum espresso, Journal of Physics: Condensed Matter 29~(46) (2017) 465901.

\bibitem{QE-2009}
P.~Giannozzi, S.~Baroni, N.~Bonini, M.~Calandra, R.~Car, C.~Cavazzoni,
  D.~Ceresoli, G.~L. Chiarotti, M.~Cococcioni, I.~Dabo, A.~{Dal Corso},
  S.~de~Gironcoli, S.~Fabris, G.~Fratesi, R.~Gebauer, U.~Gerstmann,
  C.~Gougoussis, A.~Kokalj, M.~Lazzeri, L.~Martin-Samos, N.~Marzari, F.~Mauri,
  R.~Mazzarello, S.~Paolini, A.~Pasquarello, L.~Paulatto, C.~Sbraccia,
  S.~Scandolo, G.~Sclauzero, A.~P. Seitsonen, A.~Smogunov, P.~Umari, R.~M.
  Wentzcovitch, Quantum espresso: a modular and open-source software project
  for quantum simulations of materials, Journal of Physics: Condensed Matter
  21~(39) (2009) 395502 (19pp).

\bibitem{PhysRevB.50.17953}
P.~E. Bl\"ochl, Projector augmented-wave method, Phys. Rev. B 50 (1994)
  17953--17979.
\newblock \href {https://doi.org/10.1103/PhysRevB.50.17953}
  {\path{doi:10.1103/PhysRevB.50.17953}}.

\bibitem{PhysRevB.41.7892}
D.~Vanderbilt, Soft self-consistent pseudopotentials in a generalized
  eigenvalue formalism, Phys. Rev. B 41 (1990) 7892--7895.
\newblock \href {https://doi.org/10.1103/PhysRevB.41.7892}
  {\path{doi:10.1103/PhysRevB.41.7892}}.

\bibitem{dalcorso2014}
A.~{Dal Corso}, Pseudopotentials periodic table: From h to pu, Computational
  Materials Science 95 (2014) 337--350.
\newblock \href
  {https://doi.org/https://doi.org/10.1016/j.commatsci.2014.07.043}
  {\path{doi:https://doi.org/10.1016/j.commatsci.2014.07.043}}.

\bibitem{haas2009}
P.~Haas, F.~Tran, P.~Blaha, Calculation of the lattice constant of solids with
  semilocal functionals, Phys. Rev. B 79 (2009) 085104.
\newblock \href {https://doi.org/10.1103/PhysRevB.79.085104}
  {\path{doi:10.1103/PhysRevB.79.085104}}.

\bibitem{tran2007}
F.~Tran, R.~Laskowski, P.~Blaha, K.~Schwarz, Performance on molecules,
  surfaces, and solids of the wu-cohen gga exchange-correlation energy
  functional, Phys. Rev. B 75 (2007) 115131.
\newblock \href {https://doi.org/10.1103/PhysRevB.75.115131}
  {\path{doi:10.1103/PhysRevB.75.115131}}.

\bibitem{brandes2013smithells}
E.~A. Brandes, G.~Brook, Smithells metals reference book, Elsevier, 2013.

\bibitem{liang2021}
H.~Liang, L.~Fang, S.~Guan, F.~Peng, Z.~Zhang, H.~Chen, W.~Zhang, C.~Lu,
  Insights into the bond behavior and mechanical properties of hafnium carbide
  under high pressure and high temperature, Inorganic Chemistry 60~(2) (2021)
  515--524.
\newblock \href {https://doi.org/10.1021/acs.inorgchem.0c02800}
  {\path{doi:10.1021/acs.inorgchem.0c02800}}.

\bibitem{gerward2005}
L.~Gerward, J.~{Staun Olsen}, L.~Petit, G.~Vaitheeswaran, V.~Kanchana,
  A.~Svane, Bulk modulus of ceo2 and pro2—an experimental and theoretical
  study, Journal of Alloys and Compounds 400~(1) (2005) 56--61.
\newblock \href {https://doi.org/https://doi.org/10.1016/j.jallcom.2005.04.008}
  {\path{doi:https://doi.org/10.1016/j.jallcom.2005.04.008}}.

\bibitem{kang2019}
J.~S. Kang, M.~Li, H.~Wu, H.~Nguyen, Y.~Hu, Basic physical properties of cubic
  boron arsenide, Applied Physics Letters 115~(12) (2019) 122103.
\newblock \href {https://doi.org/10.1063/1.5116025}
  {\path{doi:10.1063/1.5116025}}.

\end{thebibliography}

\end{document}